\documentclass[a4paper]{jpconf}
\usepackage[pdftex]{graphicx}
\usepackage{natbib}

\graphicspath{{FIGS/}}

\begin{document}
\title{Mass transfer dynamics in double degenerate binary systems}

\author{M. Dan, S. Rosswog and M. Br\"uggen}

\address{School of Engineering and Science, Jacobs University Bremen, Campus
  Ring 1, 28759, Bremen, Germany}

\ead{m.dan@jacobs-university.de}

% own abbreviations
\def\msun{M$_{\odot}$}
\def\Msun{M$_{\odot}$ }

\begin{abstract}
We present a numerical study of the mass transfer dynamics prior to the gravitational 
wave-driven merger of a double white dwarf system. Recently, there has been some 
discussion about the dynamics of these last stages, different methods seemed to provide
qualitatively different results. While earlier SPH simulations indicated a very quick 
disruption of the binary on roughly the orbital time scale, more recent grid-based 
calculations find long-lived mass transfer for many orbital periods. Here we demonstrate
how sensitive the dynamics of this last stage is to the exact initial conditions. We show
that, after a careful preparation of the initial conditions, the reportedly short-lived
systems undergo mass transfer for many dozens of orbits. The reported numbers of orbits
are resolution-biased and therefore represent only lower limits to what is realized in 
nature. Nevertheless, the study shows convincingly the convergence of different methods 
to very similar results.
\end{abstract}

\vspace*{-1cm}

\section{Introduction}
White dwarfs that merge under the influence of gravitational waves have been suspected
to be related to a variety of phenomena. Although not being the most popular model for 
type Ia supernovae, white dwarf mergers have despite several ups and downs defended their 
position as a serious, possible progenitor model
\cite[]{iben84,webbink84,yoon07}. If
ignited at the surface, merger remnants have been found to be transformed into an O-Ne-Mg
white dwarf that finally forms a neutron star in an accretion-induced collapse 
\cite[]{saio85,nomoto91,saio04}. Merged low-mass white dwarf mergers are
thought to
produce extreme helium stars and the majority of R Corona Borealis stars, e.g. \citet{saio02}.
A fraction of white dwarf binaries may evolve into systems that survive the 
onset of mass transfer and evolve to longer periods with ever decreasing mass 
transfer rate (e.g. \citealt{paczynski67,nelemans01}), so-called AM CVn 
systems.\\
The merger process of two white dwarfs has been modeled by a number of groups, 
\citet{benz90,rasio95,segretain97,guerrero04,yoon07}, all of the mentioned
approaches used the SPH method. Recently, grid-based 
simulations \cite[]{dsouza06,motl07} were performed where the authors
carefully
tried to reduce the angular momentum non-conservation due to advection errors 
that often plagues grid-based codes, see e.g. \citet{new97}. Their results are very 
different from most previous SPH simulations in the sense that their donor stars
are not destroyed on a dynamical time scale, but instead show long-lived
mass transfer over many orbital periods. These results have sparked some discussion
about the stability of mass transfer in double degenerate systems during 
these stages, see e.g. \citet{fryer08}.\\
In the following, we will demonstrate the sensitivity of the simulation dynamics 
on the exact initial conditions and show that with accurately constructed starting
configurations long-lived mass transfer ensues.

\vspace*{-0.3cm}
\section{Model}\label{sec:model}

\subsection{Numerical model}
We use the  smoothed particle hydrodynamics method to solve the 3D equations of fluid dynamics. 
Our code is documented in \cite{rosswog2008}. 
The system of fluid equations is closed by the HELMHOLTZ equation of
state \cite[]{timmes00}. 
It accepts an externally calculated nuclear
composition and allows a convenient coupling to a nuclear reaction network.
We use a minimal nuclear reaction network \cite[]{hix98} to determine the evolution 
of the nuclear composition and to include the energetic feedback 
onto the gas from the nuclear reactions. 
A set of only seven abundance groups 
greatly reduces the computational burden, but still reproduces the energy generation of 
all burning stages from He burning to NSE accurately. 
We use a binary tree  \cite[]{benz90} 
to search for the neighbor particles and to calculate the  gravitational
forces.

\vspace*{-0.3cm}
\subsection{Initial conditions}
\begin{figure}[!h]
  \begin{center}
    \begin{tabular}{cc}
      \vspace{-1cm} & \\
      \includegraphics[height=1.3in]{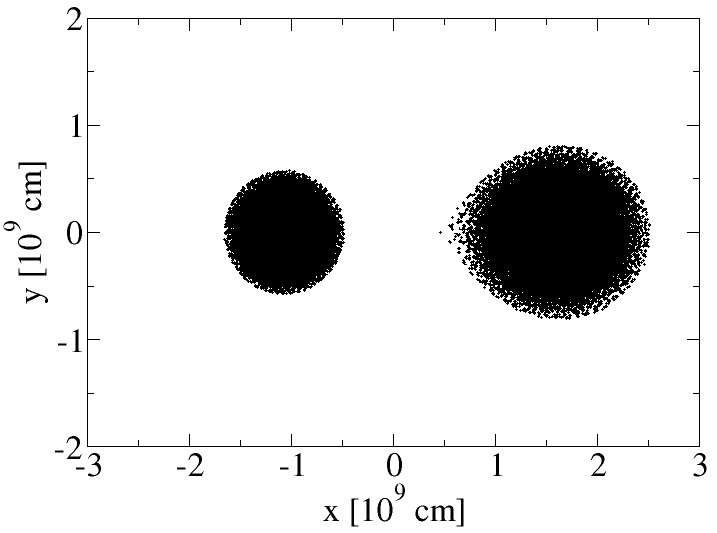} &
      \includegraphics[height=1.3in]{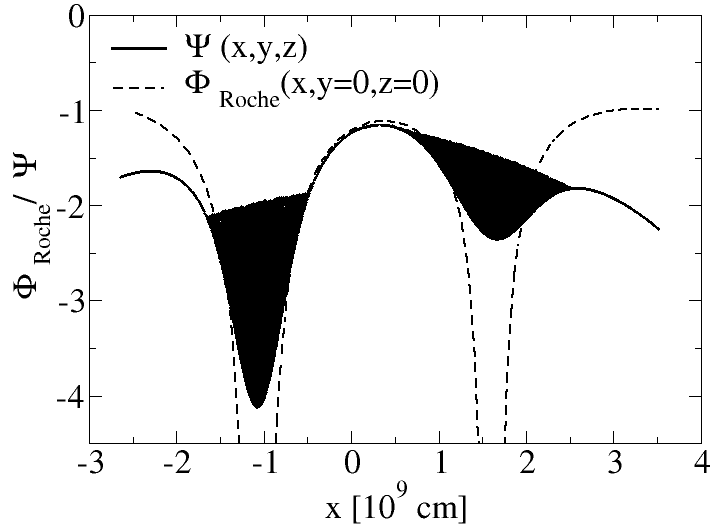}
    \end{tabular}
  \end{center}
  \caption{Equilibrium configuration ($0.9$ and $0.6\ {\rm
      M}_\odot$) at onset 
    of mass transfer. {\em Left}: projection of SPH particles onto the orbital
    plane, {\em right}: projection onto the $(x,\Phi_{\rm Roche}/\Psi)$ plane.
    The dashed line is the Roche potential as calculated for a point mass binary,
    $\Phi_{{\rm Roche}}(x,y=0,z=0)$. The solid black line is the corresponding
    quantity as derived from the real mass distribution, $\Psi(x,y=0,z=0)$,
    and is given by
$\Psi(\mbox{\boldmath${r}$})= \phi(\mbox{\boldmath${r}$}) - \frac
    12(\mbox{\boldmath${\omega}$}\times \mbox{\boldmath${r}$})^2$, where
    $\phi(\mbox{\boldmath${r}$})$ is the gravitational potential.}
  \label{fig:bineq}
\end{figure}

The dynamics of the binary systems is very sensitive to the initial conditions,
therefore, it is vital to start the simulations from a setup that is as accurate 
as possible. Most of the previous SPH simulations 
(e.g. \citealt{benz90,segretain97,guerrero04,yoon07}) started from rather approximate initial 
conditions and consequently found a very large initial mass transfer rate and a subsequent 
disruption of the donor star within a few orbits.\\
In this work, we restrict ourselves to corotating initial configurations, but very carefully 
construct initial conditions along the strategy outlined by \cite{rosswog04}. In a first
step the (isolated) stars are ``relaxed'' individually into hydrostatic equilibrium
by adding a velocity-proportional damping term to the momentum equation. Subsequently,
we place the stars in the corotating frame at a separation $a$, that is large enough 
to avoid immediate mass transfer. This separation is then adiabatically reduced
according to  $\dot a= -\frac{a}{\eta\cdot  \tau_{\rm dyn}}$,
so that $\frac{a}{\dot a} \gg \tau_{\rm dyn}$. Here, $\tau_{\rm dyn}$ is
the larger of the two white dwarf dynamical time scales and $\eta$ takes
values between $10$ and $30$. Once the first SPH particle crosses the $L_1$ point of 
the Roche potential, the binary is set on a circular orbit. This moment was adopted 
as the time origin of the simulation. An example of a corotating equilibrium
configuration at this stage (masses of $0.6\ {\rm M}_\odot$ and $0.9\ {\rm M}_\odot$) 
is shown in Fig. \ref{fig:bineq}, left panel. The right panel shows a comparison between
(point mass) Roche potential and the corresponding values as calculated from the real
matter distribution.\\

\section{From the onset of mass transfer to a possible disruption}
\label{sec:results}
\begin{figure}[!h]
  \begin{center}
    \begin{tabular}{@{}c@{}c@{}c@{}c@{}c@{}}
      \includegraphics[height=1.25in]{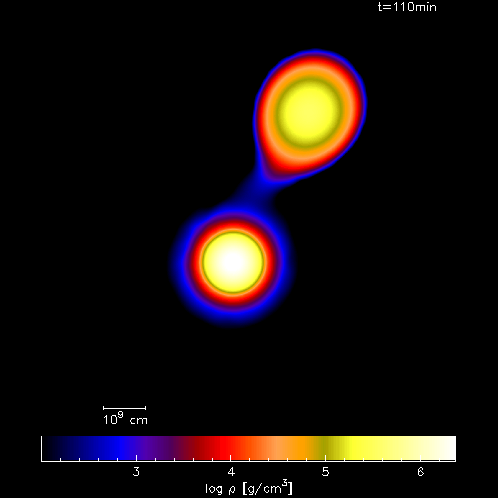} &
      \includegraphics[height=1.25in]{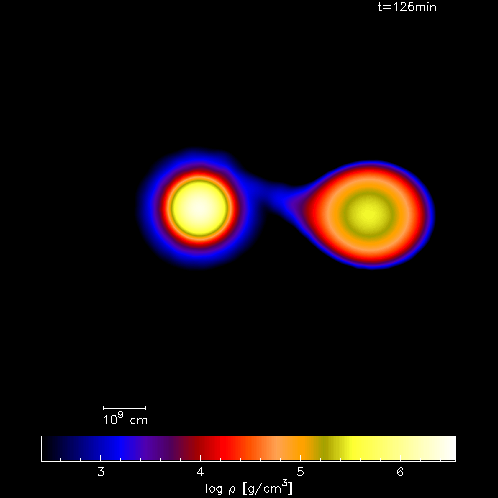} &
      \includegraphics[height=1.25in]{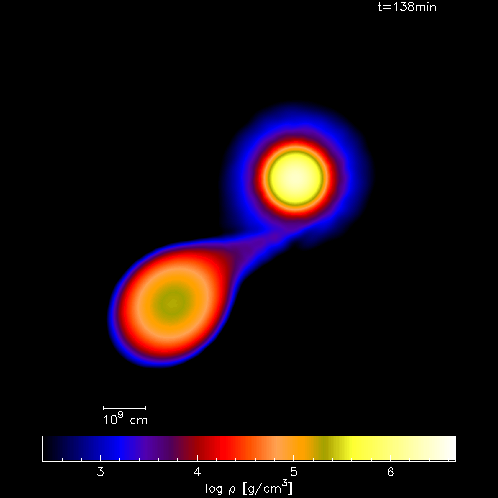}&
      \includegraphics[height=1.25in]{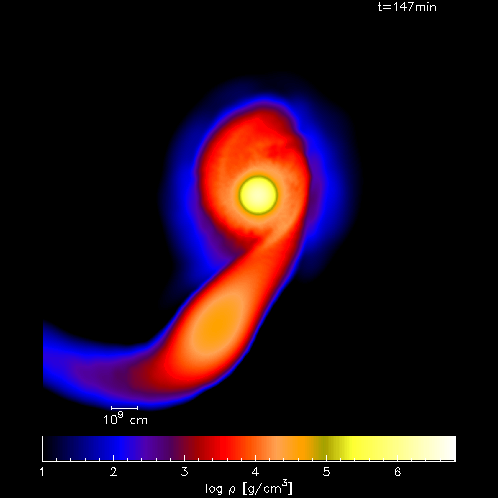} &
      \includegraphics[height=1.25in]{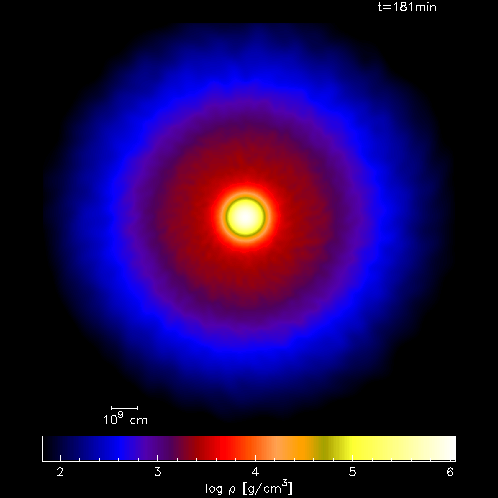} 
    \end{tabular}
  \end{center}
  \caption{Dynamical evolution of a double white dwarf system with $q=0.5$ (at
    44, 50, 56, 59 and 67 times the initial binary period.)} 
  \label{fig:WD03WD06}
\end{figure}
Exemplary for a large set of simulations that will be presented elsewhere \citep{dan09}
we show the evolution of two binary systems: one with $q= M_{\rm donor}/M_{\rm accretor}= 0.5$ 
to be compared to \cite{rasio95,dsouza06}, and another one with $q=0.27$. \cite{rasio95} also
used an SPH approach while \cite{dsouza06} used a grid-based code. Although 
we think that our investigated systems are close enough to \cite{rasio95,dsouza06} to warrant 
a fair comparison, one has to keep in mind that the latter investigations use 5/3-polytropes 
while we use a physical EOS. Since the EOS governs, via the mass-radius relationship, 
how the star reacts on mass loss, it also influences the orbital evolution. 
Thus, the comparison should be taken with a grain of salt.\\
We carry out the simulation of the $q=0.5$ case (0.3 and 0.6 \msun) at two different
resolutions. First, we use $\approx 4\times 10^4$ SPH particles to conform the 
simulations of \cite{rasio95}. We find an initial separation where (numerically 
resolvable) mass transfer sets in that is larger by 27 percent than in \cite{rasio95}. 
While in \cite{rasio95} the binary was disrupted within five orbits, we count more than 
18 full orbits before the disruption sets in.\\
In a second simulation of the same system we use $2\times 10^5$ SPH particles, see 
Fig.~\ref{fig:WD03WD06}, and consequently the initial separation increases slightly
since we are now better resolving the outer stellar layers. The main purpose of this 
run is the comparison with \cite{dsouza06} who, in turn, ran their calculation to compare 
against \cite{rasio95}. \cite{dsouza06} found the orbital separation to be constant for
about 20 orbits, then it increased and the mass transfer rate leveled off without any sign
of a merger. After 32 orbits the simulation was stopped. We followed our simulation for an 
even longer time. Similar to \cite{dsouza06} the orbital separation increases slightly but in the
end the donor is still disrupted, although only after as many as 60 orbital revolutions,
see panel 4 in  Fig.~\ref{fig:WD03WD06}. Thus, for as long as we can compare, our simulations
are in perfect agreement with \cite{dsouza06}.
\begin{figure}[!h]
      \centerline{
\includegraphics[height=1.5in]{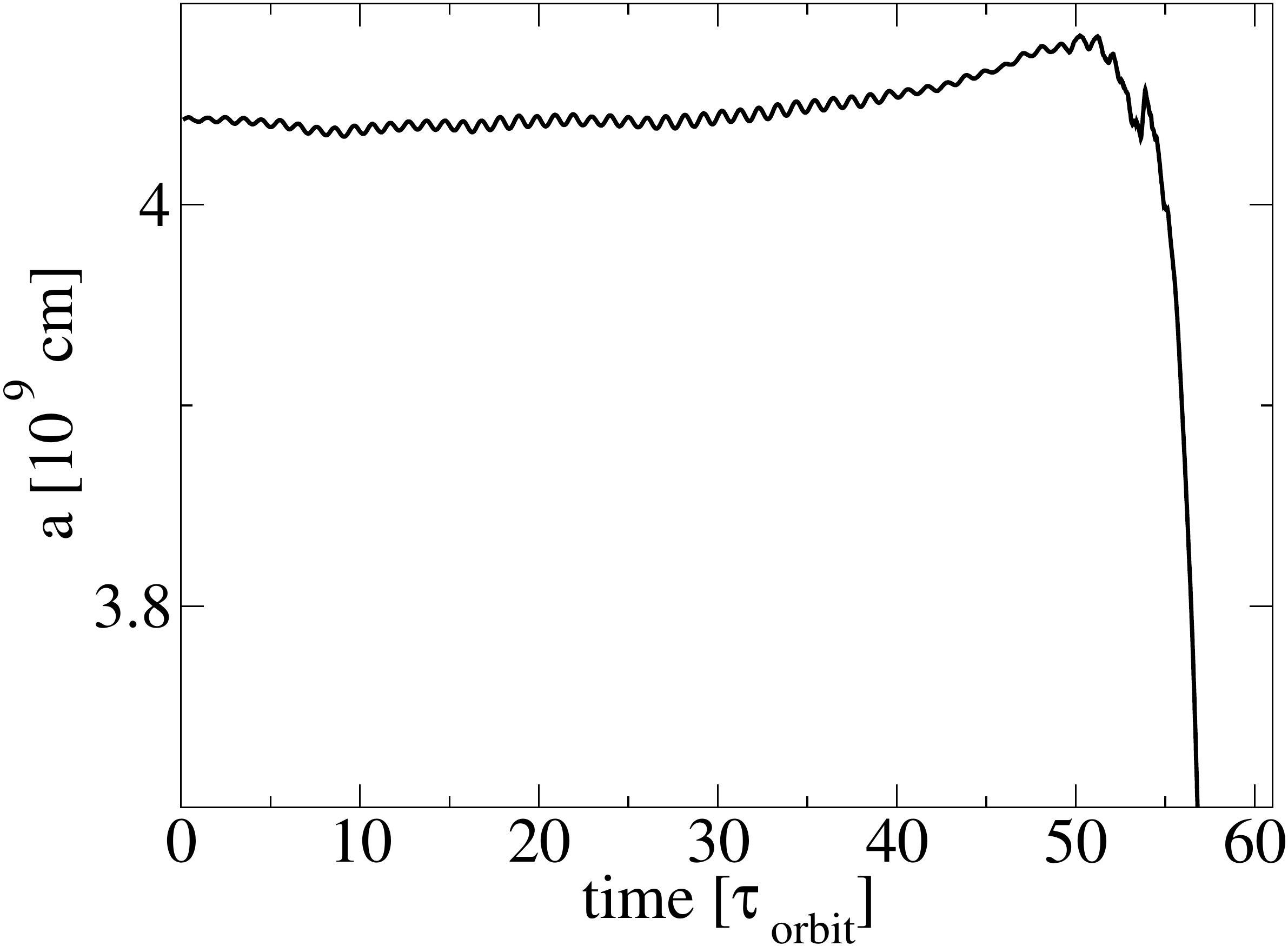}
      \includegraphics[height=1.5in]{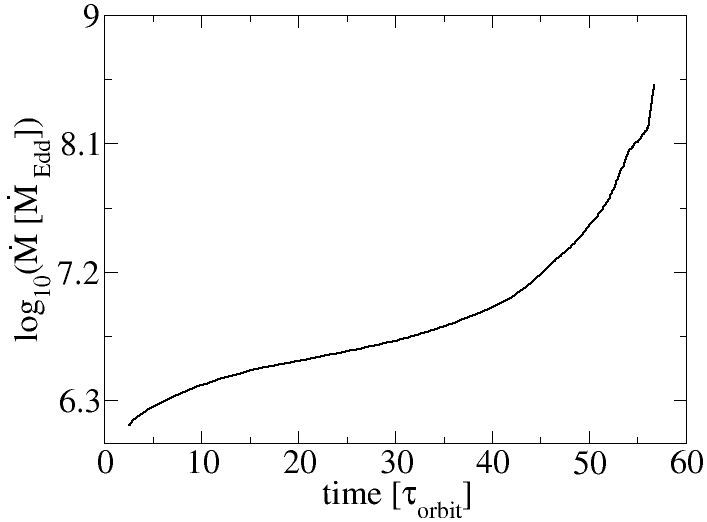}
      \includegraphics[height=1.5in]{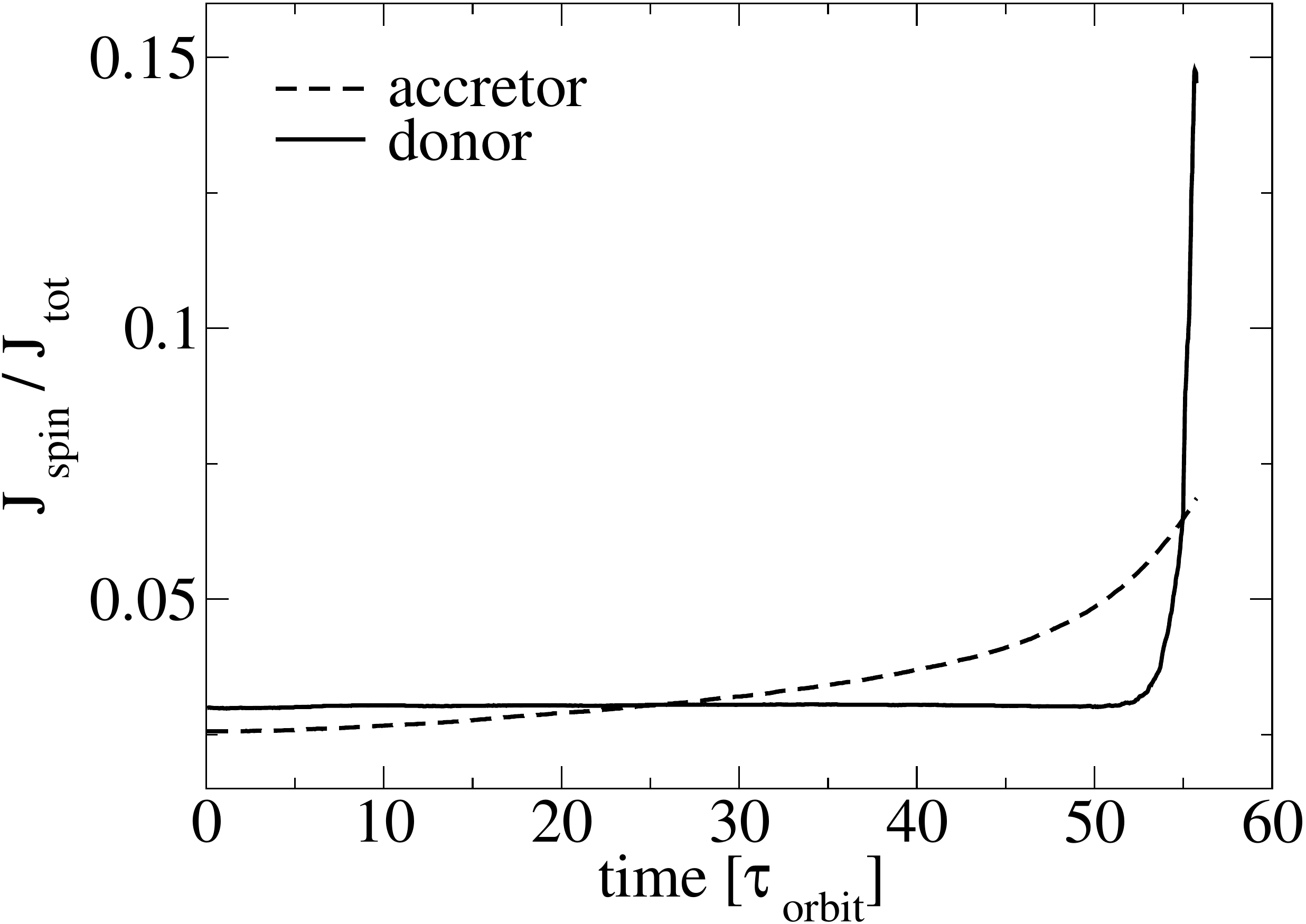}}
      \centerline{\includegraphics[height=1.5in]{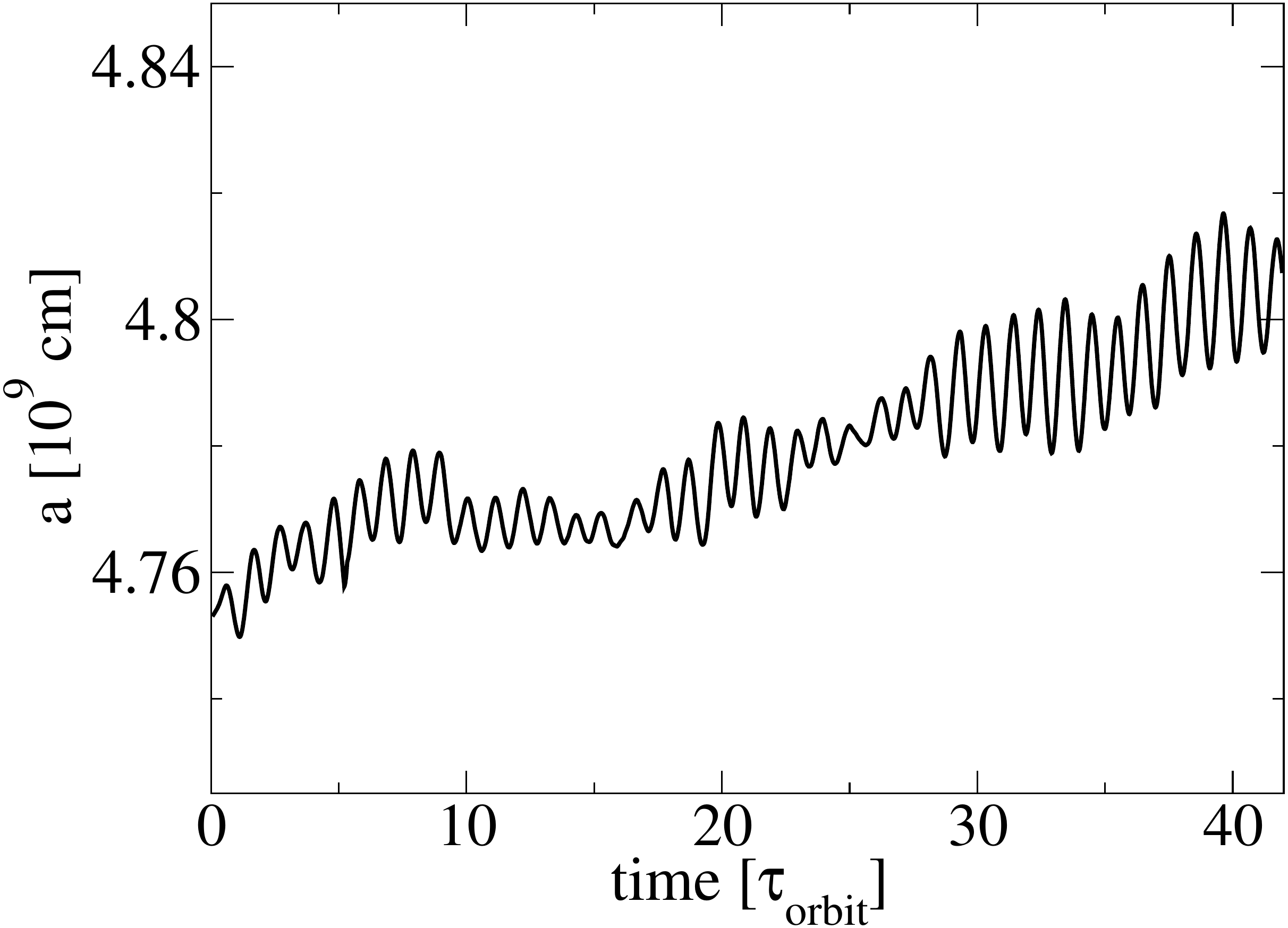}
      \includegraphics[height=1.5in]{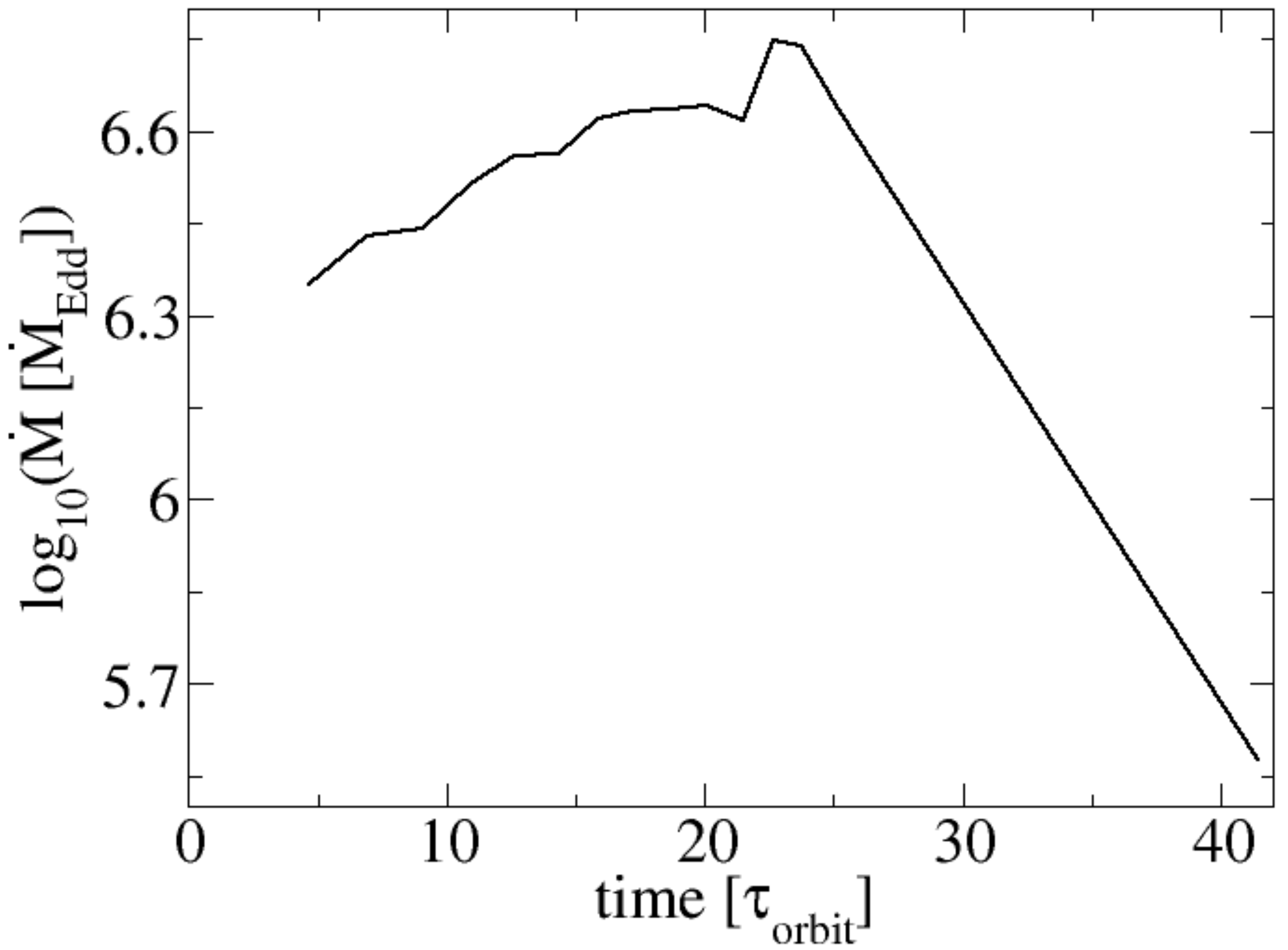}
      \includegraphics[height=1.5in]{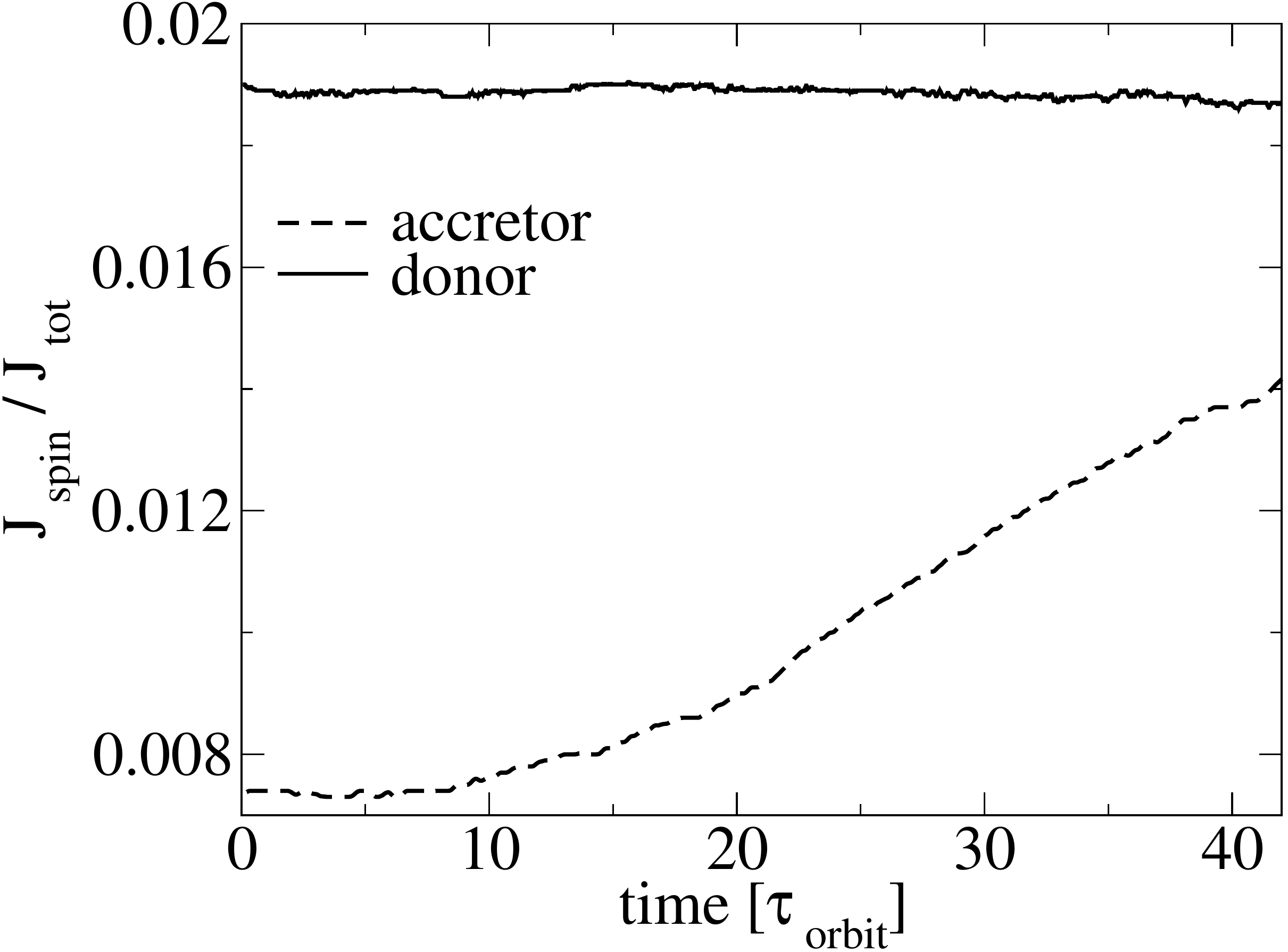}}
  \caption{Evolution of the orbital separation (left), the mass ratio (center)
    and the spin angular momentum for both the accretor and the donor (right)
    for a system with $q=0.5$ (upper panels) and $q=0.27$ (lower panels).}
  \label{fig:allquantities05_027}
\end{figure}
We also explore the evolution of a binary with a mass ration of $q=0.27$ (0.3 and
1.1 \msun). Contrary to our previous example where we saw a direct impact onto the accretor,
see Fig.~\ref{fig:WD03WD06}, panel 1, now the circularization radius is larger
than the accretor radius and a disk forms. This configuration feeds back angular momentum 
into the orbit thereby increasing the orbital separation, Fig. \ref{fig:allquantities05_027},
bottom left. This de-circularizes the orbits from the beginning. A fraction of the
angular momentum is transferred to the accretor and used to spin it up, Fig. 
\ref{fig:allquantities05_027}, bottom right. Even after 41 orbits there is no sign
of a merger, instead the separation increases continuously and the mass transfer
decreases for almost 20 full orbits (Fig. \ref{fig:allquantities05_027}, center).
Such a system could possibly evolve into a AM CVn system.

\vspace*{-0.3cm}
\section{Conclusions}
\label{sec:concl}
We have demonstrated the sensitivity to the initial conditions used in double degenerate
binary simulations and have put particular emphasis on the accuracy of our starting configurations.
Contrary to earlier SPH simulations our binaries are {\em not} disrupted after a few orbital periods,
but instead transfer mass for dozens of orbits. In as far as we can compare it, our simulations
are in perfect agreement with the recent simulations of \cite{dsouza06}.

\vspace*{-0.3cm}

\ack
This work was supported by DFG grant no. RO 3399/4-1.

\vspace*{-0.3cm}

\bibliographystyle{agsm}
\bibliography{iopart-num}

\end{document}